\documentclass[a4paper,11pt]{article}
\usepackage{pos}
\usepackage{subcaption}
\usepackage{wrapfig}
\usepackage{textpos}
\newlength{\bibitemsep}
\newlength{\bibparskip}
\let\oldthebibliography\thebibliography
\renewcommand\thebibliography[1]{%
  \oldthebibliography{#1}%
  \setlength{\parskip}{\bibitemsep}%
  \setlength{\itemsep}{\bibparskip}%
}

\title{Simulation study for the future IceCube-Gen2 surface array}
 \ShortTitle{Simulation for IceCube-Gen2 surface array}

\author{The IceCube-Gen2 Collaboration \\{\normalsize \normalfont(a complete list of authors can be found at the end of the proceedings)}}
\emailAdd{alanc@udel.edu}
\emailAdd{agnieszka.leszczynska@kit.edu}
\emailAdd{mark.weyrauch@student.kit.edu}

\abstract{The next generation of the IceCube Neutrino Observatory, IceCube-Gen2, will constitute a much larger detector, increasing the rate of high-energy neutrinos. IceCube-Gen2 will address the long-standing questions about astrophysical accelerators. The experiment will also include a surface air-shower detector which will allow for measurements of cosmic rays in the energy region where a transition between Galactic and extragalactic accelerators is expected. As a baseline design for the surface detector, we consider a surface array above the optical in-ice array consisting of the same type of stations used for the IceTop enhancement, i.e., scintillation detectors and radio antennas. In order to better understand the capabilities of such an array, we performed simulations of its response to air showers, including both detector types. We will show the results of this simulation study and discuss the prospects for the surface array of IceCube-Gen2.\\

\vspace{4mm}
{\bfseries Corresponding authors:}
Alan Coleman$^{1}$, Agnieszka Leszczy{\'n}ska$^{2*}$, Mark Weyrauch$^{2}$\\
{$^{1}$ \itshape Bartol Research Institute and University of Delaware Department of Physics \& Astronomy, Newark, DE, USA}\\
{$^{2}$ \itshape Karlsruhe Institute of Technology, Institute of Experimental Particle Physics, D-76021 Karlsruhe, Germany }\\[4mm]
$^*$ Presenter
\FullConference{37$^{\rm{th}}$ International Cosmic Ray Conference (ICRC 2021)\\
		July 12th -- 23rd, 2021\\
		Online -- Berlin, Germany}

}

\begin{document}
\maketitle
\section{Surface array of IceCube-Gen2}\label{sec:intro}
High-energy cosmic rays can enter the Earth's atmosphere, creating an avalanche of secondaries. From one perspective they bring us information about distant astrophysical objects. From another, the initiated extensive air showers can give us insights into particle interactions for energy scales at-and-above that of the LHC. A unique possibility to study both of these topics can be realized by the planned surface component of the next generation of the IceCube Observatory --- IceCube-Gen2~\cite{karle:2021icrc,Schroeder:2021icrc}. IceCube-Gen2 is foreseen to include three major additions to the current design of the Observatory. Firstly, the volume of in-ice optical detectors will be increased by about a factor of 8 which will increase the number of measured high-energy neutrinos and will allow us to do neutrino science in broad range of energies. Secondly, a sparse array of in-ice radio antennas will provide sensitivity to the highest energy neutrinos ($\gtrsim$30\,PeV) where the GZK flux is expected~\cite{hallmann:2021icrc}. This addition will help to determine their astrophysical origin, which can be associated with sources of ultra high-energy cosmic rays.
Finally, the observatory will be completed with a surface array, further extending the planned IceTop enhancement by a factor of $\approx$8, which is the focus of this work.
 
\begin{figure}[ht]
\centering
\includegraphics[width=0.98\linewidth]{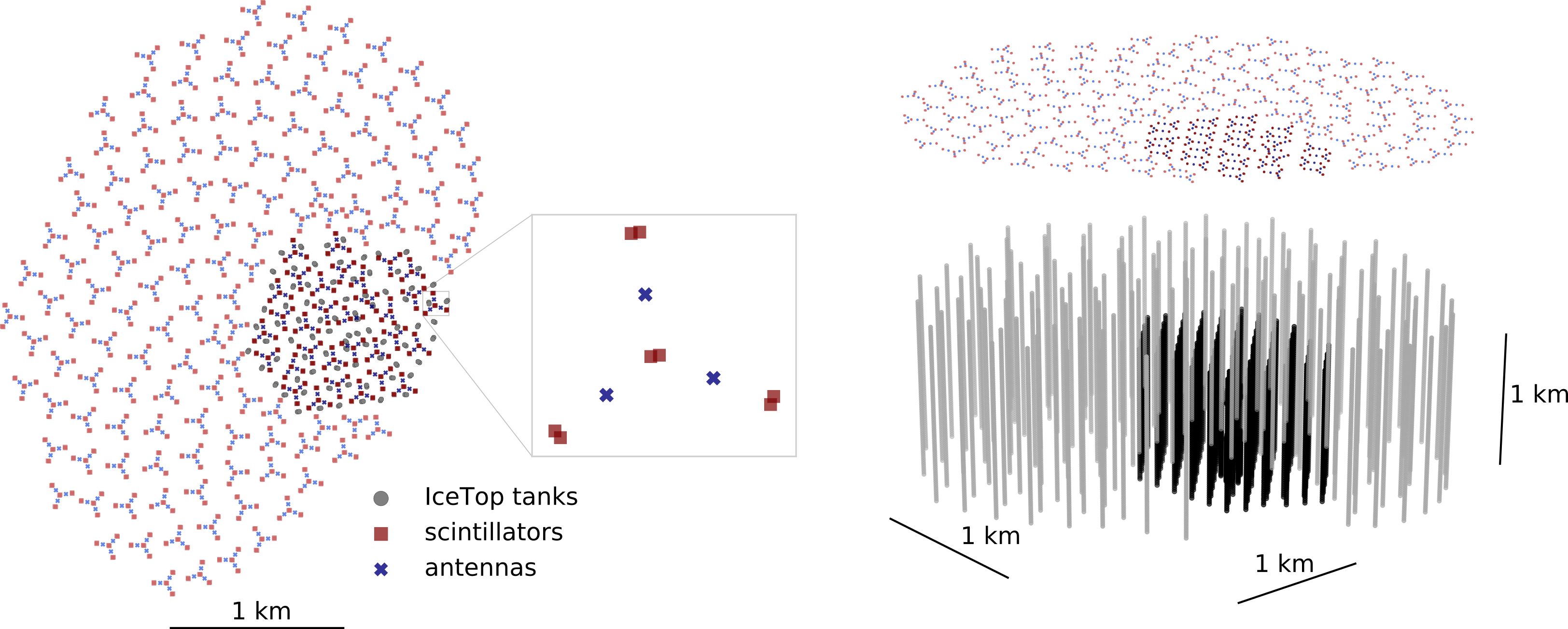}
\caption{\label{fig:layout} \textit{Left:} Surface array of IceCube-Gen2 (lighter colors) and IceTop enhancement (darker colors) consisting of hybrid stations with eight scintillation detectors and three radio antennas (inset). In addition IceTop tanks are shown. \textit{Right:} View of the optical array of IceCube-Gen2 (gray) and the current IceCube configuration (black) with the surface detectors above.}
\end{figure}

We plan to deploy, at the surface above the in-ice strings, hybrid stations consisting of eight scintillator panels (arranged in four pairs with 5\,m distance to reduce trenching length) and three radio antennas. The top view of the surface detector as well as the in-ice optical strings are shown in Figure~\ref{fig:layout}. The surface-station layout is motivated by the ongoing efforts of the IceTop enhancement~\cite{HaungsUHECR}, which consists of adding such stations within the IceTop footprint. 
The differing response of the scintillators, tanks, and radio antennas will help us to disentangle the particle content (i.e. electromagnetic and muonic) of individual air showers and increase the accuracy of reconstructing the information about the primary particle~\cite{Holt2019}.
Additionally, the array will be crucial for vetoing the air-shower particles which make up the primary background for neutrino detection in the ice.
The discrimination power will be amplified in IceCube-Gen2 due to the increase in the aperture for coincident events by a factor of $\gtrsim$30 in comparison to the existing IceCube-IceTop design. 
Indeed, the coincident measurements of secondary particles at the surface and the $\gtrsim$TeV muonic core in the in-ice array is what makes the Observatory a unique facility for studying cosmic rays \emph{and} particle physics. On the one hand, coincident measurements are a promising method in the determination of the cosmic-ray mass composition~\cite{koundal:2021icrc}. On the other hand, muon measurements depend on the simulations of extensive air showers and therefore on our understanding of the hadronic interactions. Consequently, they can contribute to the verification of the hadronic models~\cite{albrecht2021muon}.
These measurements will be enhanced by the better precision of the hybrid air-shower reconstruction.

Moreover, the larger aperture of the IceCube-Gen2 surface array will increase statistics in the measurements performed by IceTop and hence extend the energy region further into the so called \emph{transition region} where cosmic rays are expected to be dominantly accelerated by extragalactic objects. A more detailed view on different science cases which can be covered with this instrumentation are discussed in \cite{Schroeder:2021icrc}. In this work we will focus on some aspects of the capabilities of the foreseen surface array. We will present the current simulation results for the array of scintillation detectors and radio antennas.

\section{Simulation overview}\label{sec:sims}
\begin{wrapfigure}{r}{0.49\textwidth}
\vspace{-15pt}
  \begin{center}
    \includegraphics[width=0.98\linewidth]{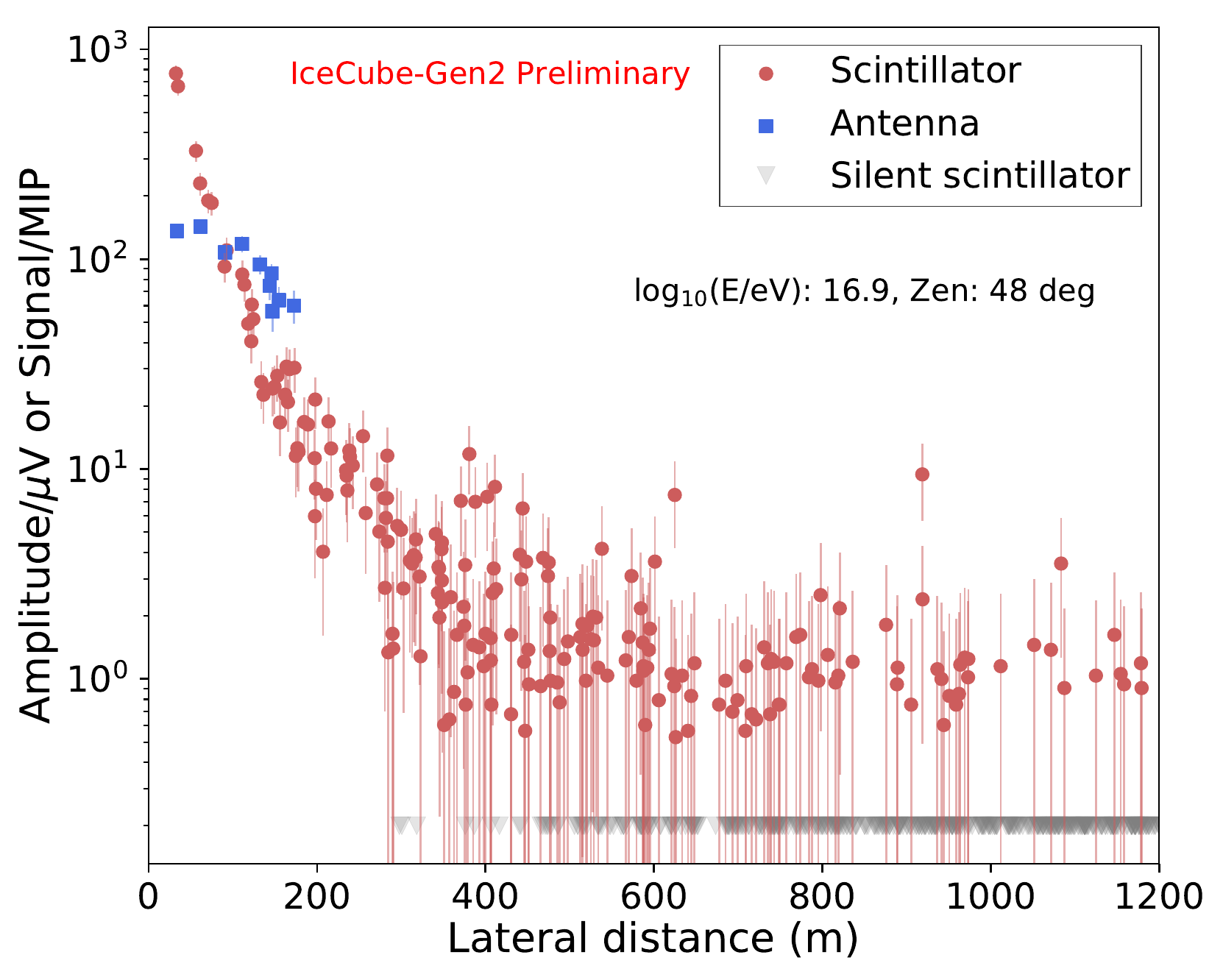}
  \end{center}
  \vspace{-11pt}
  \caption{Lateral distribution of a thinned iron shower, as observed by the scintillators (circles/triangles) and antennas (squares), is given above. There are additional hit scintillators above 1.2\,km which are not shown.}
    \label{fig:LD}
     \vspace{-5pt}
\end{wrapfigure}
To quantify the scientific potential of the design of the IceCube-Gen2 surface array, a library of air showers was produced. CORSIKA~\cite{corsika} and CoREAS~\cite{coreas} were used to simulate the surface particle content and radio emission, respectively. CORSIKA was compiled with FLUKA~\cite{Ferrari_fluka:a, Bohlen:2014buj} for low-energy interactions and Sibyll 2.3d~\cite{PhysRevD.102.063002} for high-energy ones. Due to the large computational requirements for radio simulations and to minimize the requirement to thin the simulations, two approaches were used. Simulations with only the CORSIKA routine and only the scintillator array response were performed without the thinning method for air showers with $10^{13} < E_{\rm CR} < 10^{17}$\,eV and zenith angles up to $\approx$51$^\circ$ ($\approx$72$^\circ$ for energies up to 10$^{16.5}$\,eV only for trigger efficiency studies). For the simulations which included radio emission, the thinning algorithm was applied. The external trigger from scintillation detectors was not simulated as they are no longer a limiting factor for radio antennas to trigger (Figure\,\ref{fig:trigger-scint}). Proton and iron primaries were used as limiting cases.

Each air shower is injected multiple times with cores randomly distributed over the 1.5\,km radius from the center of the IceCube-Gen2 surface array. A simulation of the detector response was then performed for both the scintillators and antennas. The secondary particles on the ground were injected into the scintillator panels and their energy losses were calculated using Geant4 toolkit (Geant4-10.4.0)\,\cite{Agostinelli:2002hh}, including a parameterization of the losses and delays in the wavelength-shifting fibres. The signal is then expressed in the units of minimum-ionizing particles (MIPs). For more details see~\cite{leszczynska2019simulation, leszczynskaPhD}. 

The radio emission from CoREAS was generated on a star-shaped pattern such that individual simulations could be re-used for such multiple core locations (see \cite{Narayan:2021icrc}). The electric field at each antenna was then convolved with the vector effective length to produce a voltage waveform. This was then further folded with the frequency-dependent response of the various amplifiers, cables, and digitizer to produce a $1\,\mu$s waveform with 1\,GHz sampling. For more see~\cite{Oehler:2021icrc,coleman:2021icrc}. Additional noise was included using the Cane model~\cite{cane1979spectra} for diffuse emission from the Galaxy. Finally, individual antennas were then selected based on a signal-to-noise ratio (SNR) cut, $(\textrm{peak}/\textrm{RMS})^2 > 18.3$.

The main cosmic ray observables are derived from the distribution of signal amplitudes and time stamps in the coordinate system of the air-shower axis. An example of such a lateral distribution of signals is shown in Figure\,\ref{fig:LD} for both detector types for an air shower initiated by an iron primary with an energy of $\approx$75\,PeV.  
For the larger IceCube-Gen2 surface footprint, the shower geometry will be very well estimated using the scintillation reconstruction. The IceTop tanks, shielded by a few meters of snow, can be used as muon-sensitive detectors at, for instance, $\gtrsim$\,600\,m from the shower axis, as in~\cite{soldin:2021icrc}. 
Regarding the radio emission, for $\theta \gtrsim 40^\circ$, a Cherenkov ring is visible\,\cite{Nelles:2014dja}, seen in the plot as a peak at about 100\,m. This structure can be used to indirectly infer the cosmic ray composition as the ring-radius is proportional to the geometric distance to X$_{\text{max}}$.

\section{Capabilities of air shower detection}\label{sec:results}
\begin{figure}[b]
\centering
\begin{subfigure}{.5\textwidth}
  \centering
  \includegraphics[width=0.98\linewidth]{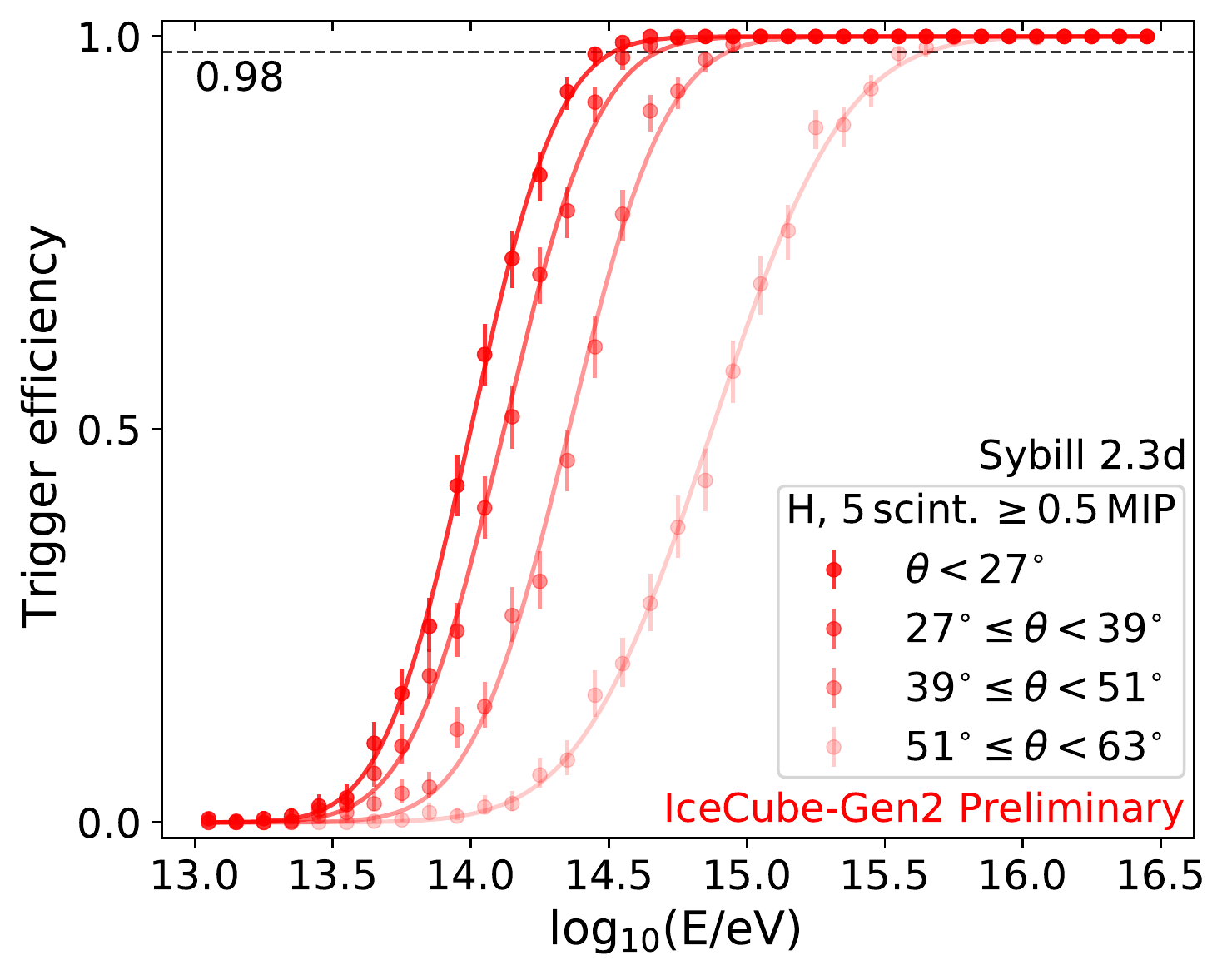}
\end{subfigure}%
\begin{subfigure}{.5\textwidth}
  \centering
  \includegraphics[width=0.98\linewidth]{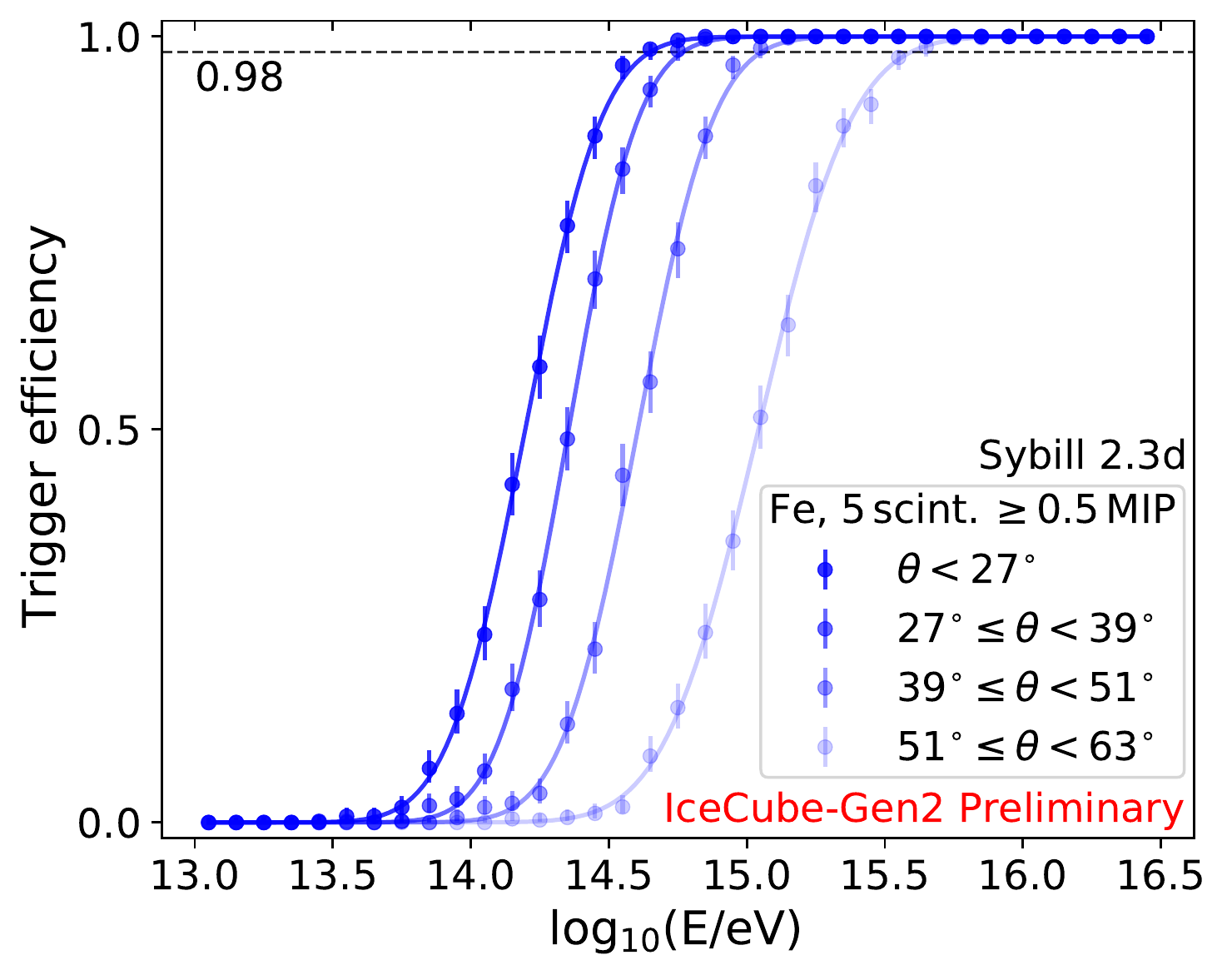}
\end{subfigure}
\caption{Trigger efficiency for the scintillator array for proton- (left) and iron- (right) induced air showers. An event is considered to be detected when at least five scintillator panels are triggered in the whole array. Each panel is triggered when a signal of at least 0.5\,MIP is obtained.}
\label{fig:trigger-scint}
\end{figure}

We foresee that the surface array will share a global triggering system between all future and existing detector components. As the scintillators are the most sensitive component of the surface hardware, they set the energy threshold at which cosmic ray detection will be possible. The efficiency curves for the scintillators, are shown in Figure\,\ref{fig:trigger-scint} for proton and iron primaries in various zenith angle bins. To show the capabilities of the whole IceCube-Gen2 surface array, a selection was performed in which true core locations were selected out to 100\,m inside the array's edge and include the current IceTop footprint. 
The scintillator array is fully efficient in triggering on $\geq$0.5\,PeV for the most vertical bin studied, but the thresholds increase with inclination mainly due to the increased absorption of the electromagnetic component with slant depth. A similar effect is visible for both species but proton-initiated showers are more efficiently detected, likely due to their deeper interaction point in the atmosphere and therefore increased particle content on the ground. 

\begin{wrapfigure}{r}{0.5\textwidth}
\vspace{-15pt}
  \begin{center}
    \includegraphics[width=0.46\textwidth]{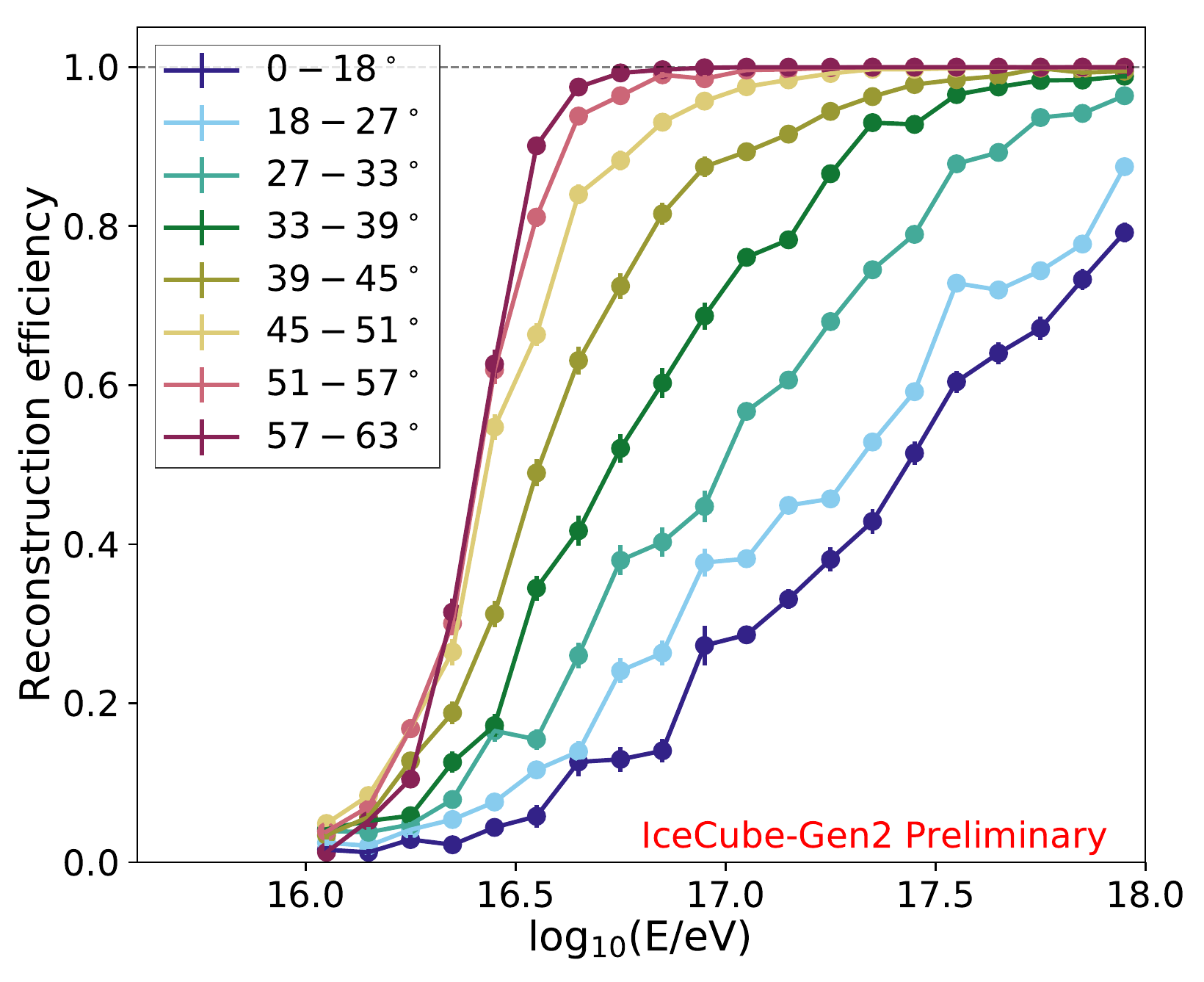}
  \end{center}
  \vspace{-11pt}
  \caption{Reconstruction efficiency for radio surface array for proton and iron induced air showers for different zenith ranges. An event is counted as reconstructed when a plane wave fit agrees to within 5$^{\circ}$ of the simulated one and at least three antennas are triggered.}
    \label{fig:radio-reco}
     \vspace{-5mm}
\end{wrapfigure} 
For showers up to 63$^{\circ}$ a threshold close to the current threshold for the IceTop high-level analysis is obtained~\cite{IceCube:2019hmk}. However, the IceTop threshold constantly increases due to the snow accumulation. 
The radio array is foreseen to be externally triggered by the other detector components. Thus we instead calculate the reconstruction efficiency by requiring that a plane-wave fit to the selected antennas (SNR > 18.3) is consistent to within 5$^\circ$ of the simulated shower axis. Contrary to the case of the scintillators, the efficiency improves with increasing zenith angle as presented in Figure~\ref{fig:radio-reco}. For quasi-vertical showers, only partial efficiency in the explored energy range is achievable, while for more inclined trajectories, the array is fully efficient near 50\,PeV.

\vspace{5mm}
\subsection{Outlook of the air shower reconstruction using scintillators}

The scintillator reconstruction is based on the efforts carried out for the IceTop enhancement and was tuned to the zenith range up to 45$^{\circ}$. The signal and timing information is included in a three-step negative-log-likelihood minimization procedure in which the shower axis and the impact point on the ground are estimated, as well as the parameters of the lateral distribution (see Figure\,\ref{fig:LD}) and shower front functions. The likelihood function takes into account the signal fluctuations obtained from the simulations of air showers, however, in the future, observational values could be used. The reconstruction of the radio waveforms beyond a simple geometric fit is currently under development. 

\begin{figure}[t]
\begin{minipage}[t]{.49\textwidth}
\centering
  \includegraphics[height=4.8cm]{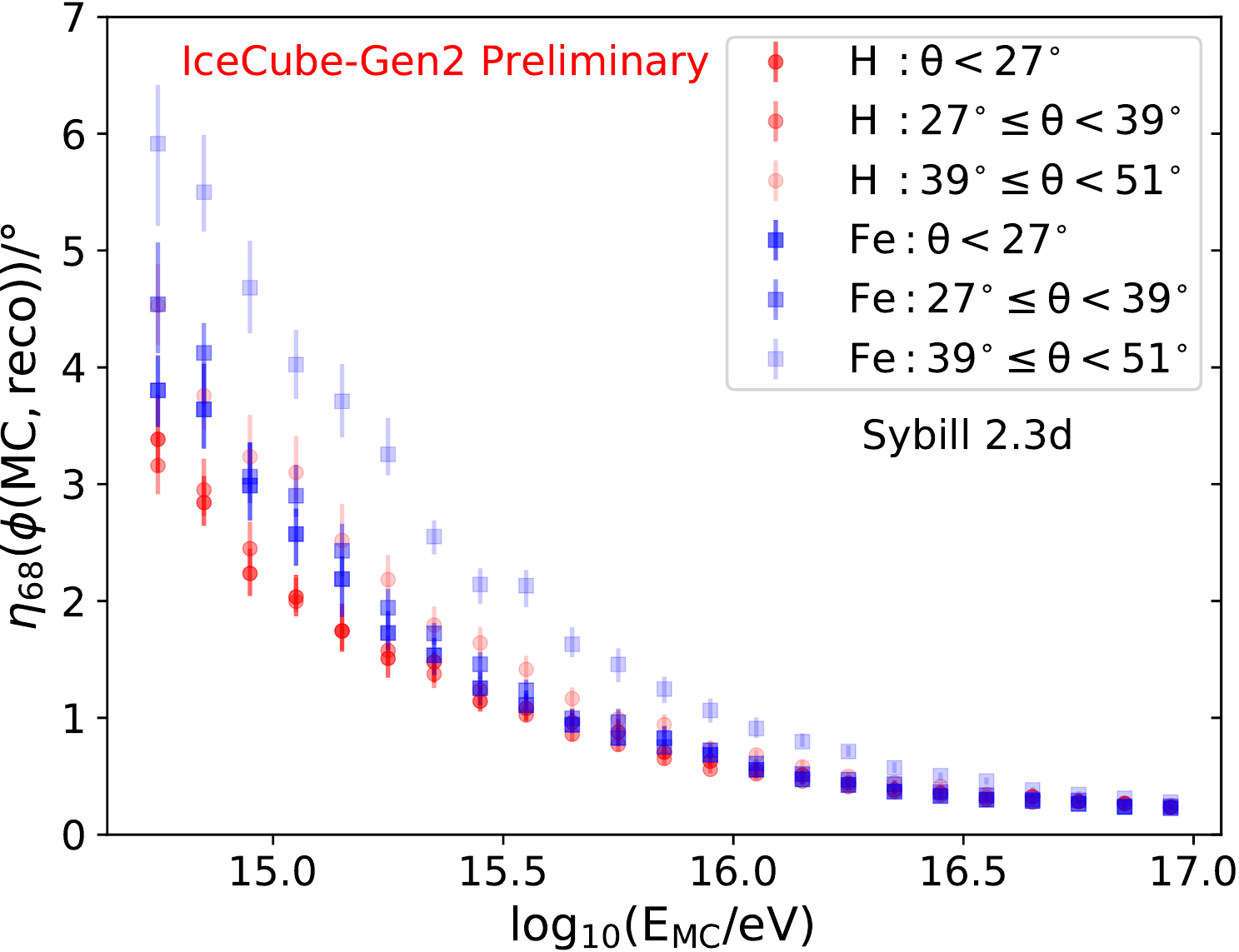}
  \caption{Angular resolution of the scintillator reconstruction for proton and iron initiated air showers for different zenith ranges. }
  \label{fig:angular}
\end{minipage}%
\hfill  
\begin{minipage}[t]{.49\textwidth}
\centering
  \includegraphics[height=4.8cm]{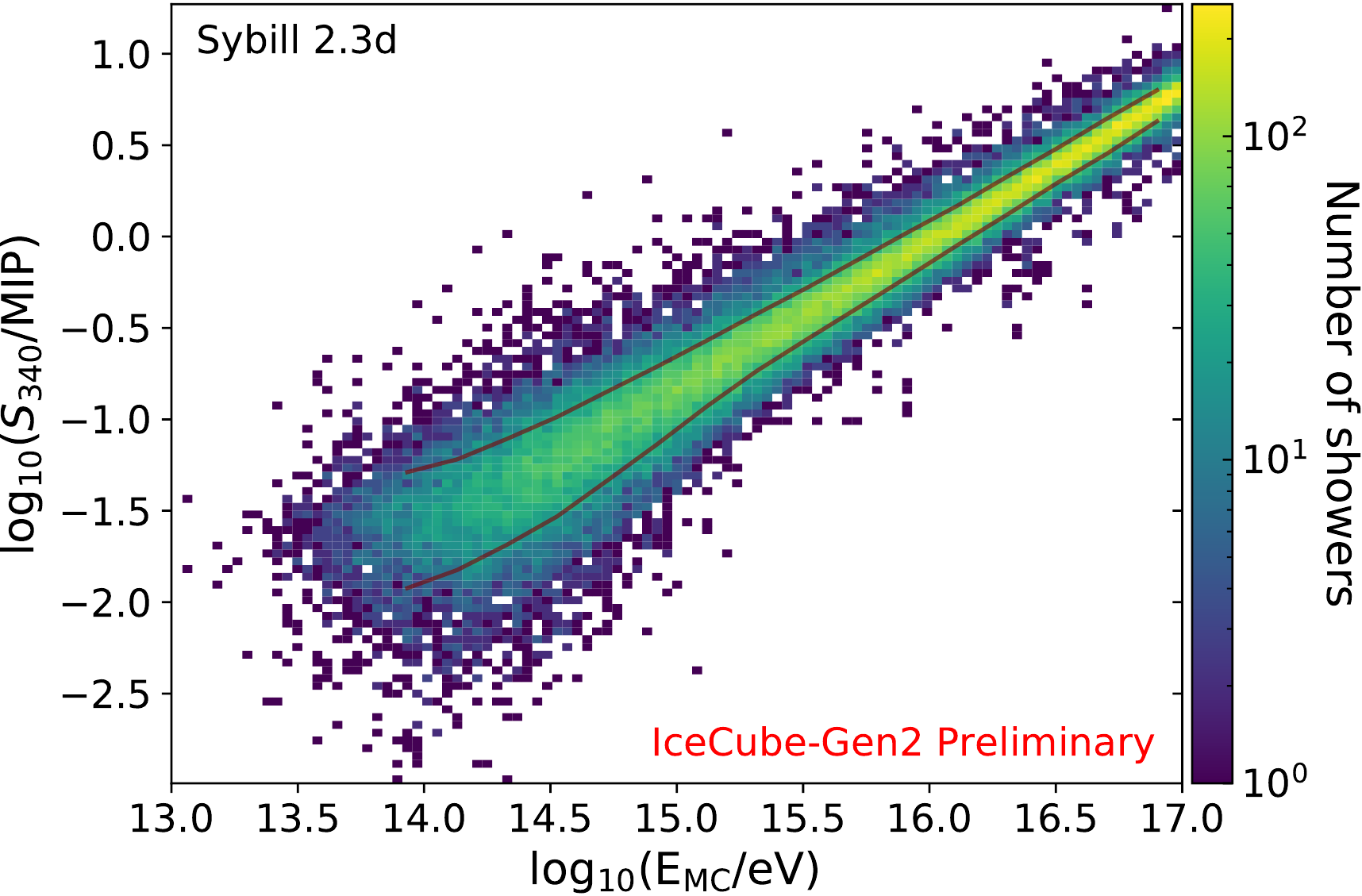}
  \caption{Distribution of the reconstructed energy estimator as a function of simulated energy for proton-induced air showers up to 45$^{\circ}$ zenith angles. Brown lines indicate 2-sigma width.}
  \label{fig:sref}
\end{minipage}%
\end{figure}

For contained showers, the resolution of the reconstructed arrival direction is shown in Figure\,\ref{fig:angular}. At the threshold energies of around 1\,PeV, the direction can be reconstructed with an accuracy of a few degrees, while above 10\,PeV, it reaches the sub-degree level. Such good estimation of the shower geometry, together with a good estimation of the impact point at the ground, will constitute a relevant input to the radio reconstruction. Further, a reliable direction estimation can be of interest for studies of cosmic-ray anisotropy. 

Preliminary studies were performed on the cosmic-ray energy estimator which was considered here as the signal expected at a lateral distance of 340\,m. In this work, the reference distance is obtained as an average from the distribution of lateral distances of triggered detectors~\cite{Kislat2012}. With this approach, a small mass dependency was achieved~\cite{leszczynska2019simulation} while still keeping good statistical precision

\noindent of the energy. The distribution for proton-initiated air showers for inclinations up to 45$^{\circ}$ is shown in Figure\,\ref{fig:sref}. The width of the distribution reflects the resolution with which one can reconstruct cosmic ray energies. It narrows towards higher energies, leading to a width of less than 0.1 in decimal logarithm of energy. Further optimisation and validation of the shower-size reconstruction algorithms are ongoing.

While radio antennas will bring an even better estimation of the electromagnetic content and therefore the energy, the scintillator array will be able to extend the measurements beyond the second knee at 100\,PeV and make comprehensive studies across several decades in energy. Moreover, radio emission provides a measurement of X$_{\text{max}}$, which is sensitive to cosmic ray mass.

\section{Conclusions and future prospects}\label{sec:future}
The IceCube Observatory with its surface detectors (and enhancement) is unique for studying cosmic rays as it allows for the study of the many low-energy muons and electromagnetic particles while simultaneously measuring $\approx$\,TeV muons in the in-ice volume. Combining the observations in the ice with the IceTop measurements has already proven useful for understanding the cosmic-ray mass distribution~\cite{IceCube:2019hmk} and discerning between hadronic interaction models~\cite{Verpoest:2021icrc}.

The planned IceCube-Gen2 can enhance cosmic ray studies by extending the phase space to higher angles and higher energies, with an increase by a factor of $\gtrsim$30 in geometric aperture for coincident events compared to IceCube. Further, increased discrimination power will come from the inclusion of the X$_\textrm{max}$ measurements by the radio antennas and the calorimetric energy measurements, particularly in the transition region between Galactic and extragalactic sources.

\begin{wrapfigure}{r}{0.49\textwidth}
    \centering 
    \includegraphics[width=0.97\linewidth]{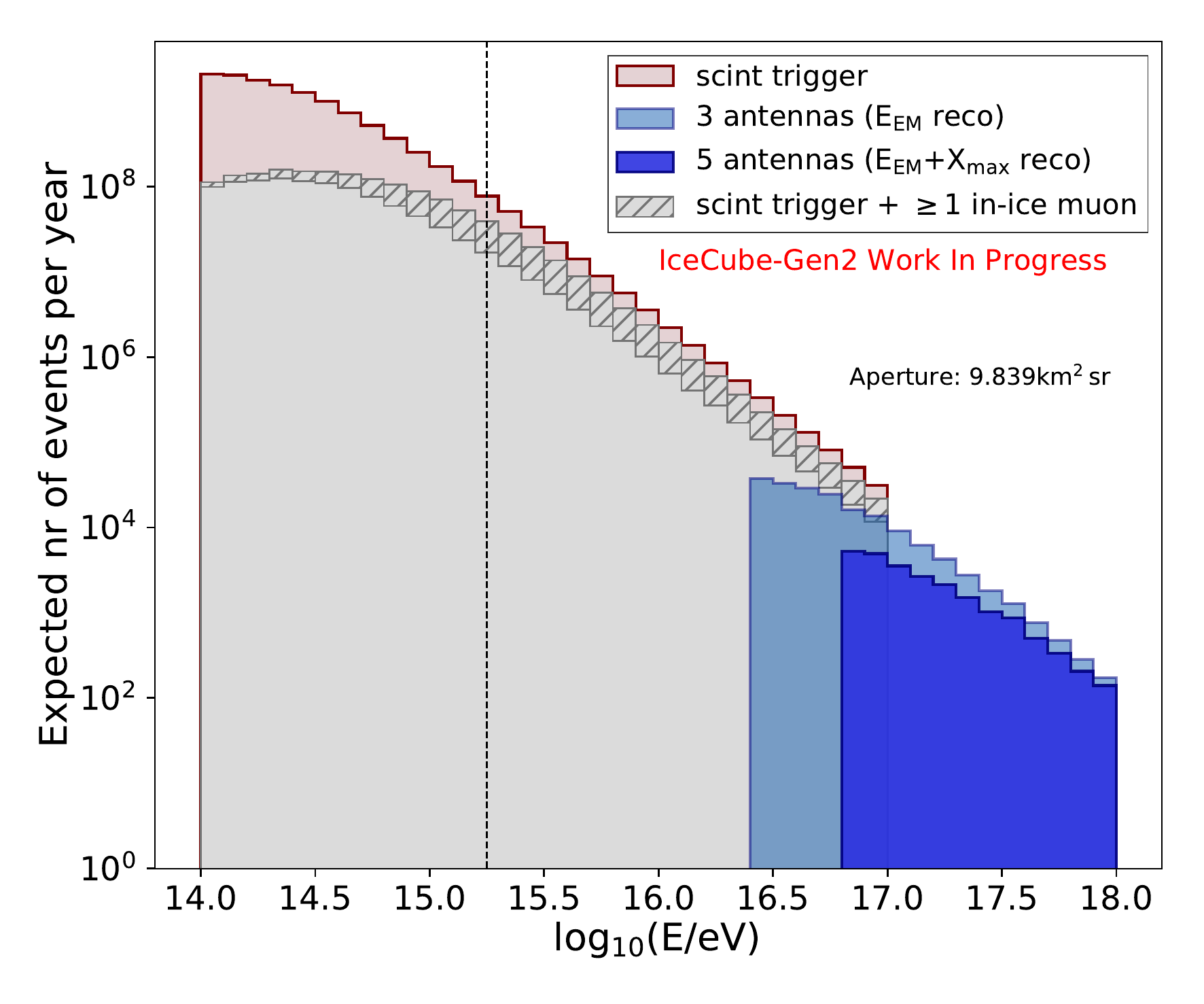}
  \vspace{-5pt}
  \caption{Event rate for CRs with $\theta\lesssim51^\circ$ assuming an H4a flux for scintillator-triggered events (red), scintillator trigger and a reconstructed in-ice muon(s) (gray). The rates for which the EM energy content (blue) as well as X$_\textrm{max}$ (dark blue) of a shower can be determined using the antennas are shown. Vertical dashed line indicates the energy where the scintillator array reaches full trigger efficiency ($\theta\lesssim51^\circ$).}
    \label{fig:rates}
\end{wrapfigure}

This net effect of the increase in effective area is shown in Figure\,\ref{fig:rates} by the expected number of events per year which is obtained for various detection schemes. The cosmic ray flux, given by the H4a model~\cite{Gaisser:2013bla}, was multiplied by the aperture (containment area of the IceCube-Gen2 surface array including IceTop and its enhancement footprint, zenith range up to $\approx$51$^{\circ}$) and a given detection probability. The various probabilities were calculated using the number of hit scintillators, the observed  
antenna signals, and after propagating the muons through the in-ice volume\,\cite{KOEHNE20132070}, as needed. For the scintillator array, we require at least 5 hit detectors within the full array. For the radio array, we \emph{additionally} require that there are at least 3 antennas with signals with SNR $>$ 18.3 to reconstruct the electromagnetic energy content, $E_{\rm EM}$, of the shower.
A requirement of at least 5 antennas is used as a proxy for the events for which we may be able to reconstruct $X_{\rm max}$ as well~\cite{Bezyazeekov:2018yjw}.
The events with an in-ice muon require a scintillator trigger and that the event can be reconstructed by the optical detectors.
In the absence of a full in-ice reconstruction, we instead calculate an upper and lower limit given by an optimistic and pessimistic situation.
We use the probability that we can use the optical detector to reconstruct a single muon with energy (see \cite{omeliukh:2021icrc} for details), $E_\mu$, and zenith angle, $\theta$, at the edge of the in-ice volume, $P(E_{\mu}, \theta)$. We then calculate, for individual events,
\begin{enumerate}
    \item the probability to reconstruct at least one of $N$ uncorrelated muons, $1 - \prod\limits_i [1 - P(E_{\mu,i}, \theta_i)]$,
    \item the probabilty to reconstruct the most energetic muon, $P(E_{\mu, max}, \theta)$,
\end{enumerate}
given the $N$ muons that propagate to the edge of the detector volume.
Near the second knee, the array will detect $\mathcal{O}(10^{2} - 10^{4})$ events per year with many including a reconstruction of $X_{\rm max}$, $E_{\rm EM}$, and/or the $\sim$1\,GeV (surface) and $\sim$1\,TeV (in-ice) muon content. 

With these capabilities the new detector will complement the current cosmic-ray measurements from other experiments. At lower energies, the measurement will overlap with, for example, LHAASO\,\cite{DISCIASCIO2016166} which studies cosmic rays from Galactic sources, with an order of magnitude higher statistics provided by the IceCube-Gen2 surface array. It will also allow for an overlap at higher energies with the low-energy extensions of Auger\,\cite{Coleman:2020nqn} and TA\,\cite{Abbasi_2021}. Thus, the IceCube-Gen2 surface array will play an important part in understanding the transition between Galactic and extra-galactic sources. The multi-hybrid measurements at the surface and in the ice provide a unique advantage, not only for air shower analyses but also for astrophysical neutrino searches.

{\small
\ \\
    \noindent\textbf{Acknowledgement:}
    The authors acknowledge support by the High Performance and Cloud Computing Group at the Zentrum für Datenverarbeitung of the University of Tübingen, the state of Baden-Württemberg through bwHPC and the German Research Foundation through grant no INST 37/935-1 FUGG. Additional support also from U.S. National Science Foundation-EPSCoR (RII Track-2 FEC, award ID 2019597).
}
% \newpage

\bibliographystyle{ICRC}
\bibliography{references}

% Full authors list (ONLY FOR COLLABORATIONS)
\clearpage
\section*{Full Author List: IceCube-Gen2 Collaboration}

% \noindent \textbf{Note comment afterwards:} Collaborations have the possibility to provide an authors list in xml format which will be used while generating the DOI entries making the full authors list searchable in databases like Inspire HEP. For instructions please go to icrc2021.desy.de/proceedings or contact us under icrc2021proc@desy.de.\\

% \scriptsize
% \noindent
% first.author$^1$, 
% second.author$^2$, 
% third.author$^3$ % .... more names
% and 
% last.author$^{n}$ \\

% \noindent
% $^1$first.affiliation.
% $^2$second.affiliation. % .... more affiliation
% $^{m}$last.affiliation.

\scriptsize
\noindent
R. Abbasi$^{17}$,
M. Ackermann$^{71}$,
J. Adams$^{22}$,
J. A. Aguilar$^{12}$,
M. Ahlers$^{26}$,
M. Ahrens$^{60}$,
C. Alispach$^{32}$,
P. Allison$^{24,\: 25}$,
A. A. Alves Jr.$^{35}$,
N. M. Amin$^{50}$
% ...
\\
\\
% ...
$^{12}$ Universit{\'e} Libre de Bruxelles, Science Faculty CP230, B-1050 Brussels, Belgium \\
% ...
$^{17}$ Department of Physics, Loyola University Chicago, Chicago, IL 60660, USA \\
% ...
$^{22}$ Dept. of Physics and Astronomy, University of Canterbury, Private Bag 4800, Christchurch, New Zealand \\
% ...
$^{24}$ Dept. of Astronomy, Ohio State University, Columbus, OH 43210, USA \\
$^{25}$ Dept. of Physics and Center for Cosmology and Astro-Particle Physics, Ohio State University, Columbus, OH 43210, USA \\
$^{26}$ Niels Bohr Institute, University of Copenhagen, DK-2100 Copenhagen, Denmark \\
% ...
$^{32}$ D{\'e}partement de physique nucl{\'e}aire et corpusculaire, Universit{\'e} de Gen{\`e}ve, CH-1211 Gen{\`e}ve, Switzerland \\
% ...
$^{35}$ Karlsruhe Institute of Technology, Institute for Astroparticle Physics, D-76021 Karlsruhe, Germany  \\
% ...
$^{50}$ Bartol Research Institute and Dept. of Physics and Astronomy, University of Delaware, Newark, DE 19716, USA \\
% ...
$^{60}$ Oskar Klein Centre and Dept. of Physics, Stockholm University, SE-10691 Stockholm, Sweden \\
% ...
$^{71}$ DESY, D-15738 Zeuthen, Germany \\
% ...

\subsection*{Acknowledgements}

\noindent
USA {\textendash} U.S. National Science Foundation-Office of Polar Programs,
U.S. National Science Foundation-Physics Division,
U.S. National Science Foundation-EPSCoR,
Wisconsin Alumni Research Foundation,
Center for High Throughput Computing (CHTC) at the University of Wisconsin{\textendash}Madison,
Open Science Grid (OSG),
Extreme Science and Engineering Discovery Environment (XSEDE),
Frontera computing project at the Texas Advanced Computing Center,
U.S. Department of Energy-National Energy Research Scientific Computing Center,
Particle astrophysics research computing center at the University of Maryland,
Institute for Cyber-Enabled Research at Michigan State University,
and Astroparticle physics computational facility at Marquette University;
Belgium {\textendash} Funds for Scientific Research (FRS-FNRS and FWO),
FWO Odysseus and Big Science programmes,
and Belgian Federal Science Policy Office (Belspo);
Germany {\textendash} Bundesministerium f{\"u}r Bildung und Forschung (BMBF),
Deutsche Forschungsgemeinschaft (DFG),
Helmholtz Alliance for Astroparticle Physics (HAP),
Initiative and Networking Fund of the Helmholtz Association,
Deutsches Elektronen Synchrotron (DESY),
and High Performance Computing cluster of the RWTH Aachen;
Sweden {\textendash} Swedish Research Council,
Swedish Polar Research Secretariat,
Swedish National Infrastructure for Computing (SNIC),
and Knut and Alice Wallenberg Foundation;
Australia {\textendash} Australian Research Council;
Canada {\textendash} Natural Sciences and Engineering Research Council of Canada,
Calcul Qu{\'e}bec, Compute Ontario, Canada Foundation for Innovation, WestGrid, and Compute Canada;
Denmark {\textendash} Villum Fonden and Carlsberg Foundation;
New Zealand {\textendash} Marsden Fund;
Japan {\textendash} Japan Society for Promotion of Science (JSPS)
and Institute for Global Prominent Research (IGPR) of Chiba University;
Korea {\textendash} National Research Foundation of Korea (NRF);
Switzerland {\textendash} Swiss National Science Foundation (SNSF);
United Kingdom {\textendash} Department of Physics, University of Oxford.

\end{document}